%Paper: gr-qc/9403004
%From: BORDE@BNLCL6.BNL.GOV
%Date: Wed, 2 Mar 1994 1:51:52 -0500 (EST)

%%% The Impossibility of Steady-State Inflation
%%% Arvind Borde and Alexander Vilenkin
%%% Tufts Institute of Cosmology
%%% To appear in the proceedings of the Eighth Yukawa Symposium.
%%%
%%% IMPORTANT PROCESSING NOTE:
%%%       Use Plain TeX to `tex' this file, not LaTeX.
%%%       The resulting dvi file will have blank spaces in place of the
%%%       figures.
%%%       If you have a PostScript printer, *and* you use Tom Rokicki's
%%%       TeX-to-PostScript converter, dvips, then the figures will
%%%       also show up (after `texing' the file, say `dvips filename', then
%%%       print the PostScript file that emerges).
%%%
%%%       The original article was produced with the PostScript versions
%%%       of Times-Roman and Avante Garde typefaces, along with
%%%       Michael Spivak's MathTime mathematics typefaces. It also used
%%%       various other special commands. The present version is set up
%%%       to use the standard TeX Computer Modern typefaces and to be
%%%       self-contained.
%%%
%%%%%%%%%%%%%%%%%%%%%%%%%%%%%%%%%%%%%%%%%%%%%%%%%%%%%%%%%%%%%%%%%%%%%%%%%%%%%

\newif\ifTimesRoman % If Times-Roman is to be used.
\TimesRomanfalse

\ifTimesRoman
   \input mtmacs  % Commands for Mathematics Times faces.
   \input 8pt-tms % Commands for 8-point Times faces.

   \def\MT {Math\kern-.1emT\kern.01em\i me}
   \MTMI{10pt}{7pt}{5pt}
   \MTSY{10pt}{7pt}{5pt}
   \MTEX{10pt}
   \MathRoman{ptmr}{10pt}{7pt}{5pt}
   \MathBold{ptmb}{10pt}{7pt}{5pt}
   \font\CalFont=pzcmi at 12pt
   \def\cal #1{\ifmmode \hbox{\CalFont #1}\,\else {\CalFont #1\/}\fi}
   \font\tenrm=ptmr
    \def\rm{\fam0 \tenrm}
   \font\tenbf=ptmb
   \def\bf{\fam\bffam\tenbf}
   \font\tenit=ptmri
   \def\it{\fam\itfam\tenit}
   \font\tensl=rptmro
   \def\sl{\fam\slfam\tensl}
   \rm

   \font\vlbf=rpagd at 13pt % AvantGarde-Demi
   \font\lbf=rpagd at 10pt
   \magnification=1175
   \SmallFootnotes
\else
   \font\vlbf=cmbx10 at 12pt
   \font\lbf=cmbx10
   \let\Eightit=\tenit
   \let\Eightrm=\tenrm
   \let\Sixrm=\sevenrm
   \def\Eightpoint{}
\fi

\hsize 34 true pc
\vsize 52 true pc
\parindent 3 true pc

\def\\{\ifhmode\hfil\break\fi}
\def\footnoterule{\kern -3pt \hrule width 4 true cm \kern 2.6pt}
\footline={\hfil \lower20pt\hbox{\Eightrm $-$ \folio\ $-$}\hfil}

\def\Draft{\headline={\ifnum\pageno=1
           \vbox{\Eightit \baselineskip=2.5ex \noindent
                 \rightline{To appear in the proceedings of the}
                 \rightline{Eighth Yukawa Symposium on Relativistic
                            Cosmology, Japan, 1994.}}%
           \else \hfil \fi}}

% Referencing commands:
\newcount\RefNo \RefNo=1
\newbox\RefBox

\def\Jou #1{{\it #1}}
\def\Vol #1{{\bf #1}}
\def\AddRef #1{\setbox\RefBox=\vbox{\unvbox\RefBox
                       \parindent1.75em
                       \vskip0pt
                       \item{\the\RefNo.}
                       \frenchspacing
                       \pretolerance1000 \tolerance1000
                       \hbadness1000
                       \strut#1\strut}%
   \global\advance\RefNo by1 }
\def\Ref [#1]{~[{\the\RefNo}]\AddRef{#1}}
\def\Refc [#1]{~[{\the\RefNo,}\AddRef{#1}}
\def\Refm [#1]{$\,${\the\RefNo,}\AddRef{#1}}
\def\Refe [#1]{$\,${\the\RefNo}]\AddRef{#1}}
\def\ShowReferences {\Sectionvar{References}\par
   \unvbox\RefBox}
\def\StoreRef #1{\edef#1{\the\RefNo}}

% Sectioning commands:
\newcount \Sno \Sno=0
\def\Section #1\par{\goodbreak\vskip\baselineskip
                    \medskip
                    \global\advance\Sno by1
                    \leftline{\lbf \the\Sno.\ \ #1}
                    \nobreak\vskip.8\baselineskip
                    \nobreak\vskip-3\parskip \noindent}%
\def\Sectionvar #1\par{\goodbreak\vskip\baselineskip
                    \medskip
                    \leftline{\lbf #1}
                    \nobreak\vskip.8\baselineskip
                    \nobreak\vskip-3\parskip \noindent}%

% Equation Numbering commands:
\newcount\EqNo \EqNo=0
\def\NumbEq {\global\advance\EqNo by 1
             \eqno(\the\EqNo)}
\def\PrevEq {(\the\EqNo)}
\def\PrevEqs #1{{\count255=\EqNo \advance\count255 by-#1\relax
                 (\the\count255)}}
\def\NameEq #1{\xdef#1{(\the\EqNo)}}

\def\AppndEq{\EqNo=0
  \def\NumbEq {\global\advance\EqNo by 1
             \eqno(A\the\EqNo)}
  \def\numbeq {\global\advance\EqNo by 1
             (A\the\EqNo)}
  \def\PrevEq {(A\the\EqNo)}
  \def\PrevEqs ##1{{\count255=\EqNo \advance\count255 by-##1\relax
                 (A\the\count255)}}
  \def\NameEq ##1{\xdef##1{(A\the\EqNo)}}}

% Figure commands:
\newcount\FigNo \FigNo=0
\def\Fig {fig.~\the\FigNo}
\def\NFig {{\count255=\FigNo \advance\count255 by 1 fig.~\the\count255}}
\def\StartFigure {\advance\FigNo by 1
    \midinsert
    \removelastskip\bigskip
    \bgroup
    \dimen0=\hsize \advance\dimen0 by-2\parindent
    \indent
    \vbox\bgroup\hsize\dimen0 \noindent\hfil}
\def\caption #1{{\medskip \Eightpoint
                 \noindent{\bf Figure~\the\FigNo:} #1\par}}
\def\label (#1,#2)#3{{\offinterlineskip
                     \count1=#1 \count2=#2
                     \multiply \count1 by 1175 \divide \count1 by \mag
                     \vskip-\parskip\vbox to 0pt{\vss
                     \moveright \count1 bp \hbox to 0pt{\raise \count2 bp
                     \hbox{#3}\hss}\hskip -\count1 bp \relax}}}
\def\EndFigure {\egroup\egroup\bigskip\endinsert}

%%%%%%%%%%%%%%%%%%%%%%%%%%%%%%%%%%%%%%%%%%%%%%%%%%%%%%%%%%%%%%%%%%%%%%%%%%%%%

\Draft
\topglue 6\bigskipamount
\leftline{\indent\vlbf The Impossibility of Steady-State Inflation}
\bigskip
\bigskip
\bigskip
{\leftskip\parindent \rightskip\parindent \noindent
Arvind Borde$^{\dag\ast}$ and Alexander Vilenkin$^{\star}$
\smallskip \noindent
\sl Institute of Cosmology, Department of Physics and Astronomy,\\
Tufts University, Medford, MA 02155, USA.\par}

{\parindent1em \baselineskip12pt
\vfootnote{${\dag}$}{Permanent addresses:
Long Island University, Southampton,
N.Y. 11968, and\\
\indent High Energy Theory Group, Brookhaven National Laboratory,
Upton, N.Y. 11973.}
\vfootnote{${\ast}$}{Electronic mail: borde@bnlcl6.bnl.gov}
\vfootnote{${\star}$}{Electronic mail: avilenki@pearl.tufts.edu}
\par}

\bigskip
\bigskip
\bigskip
\bigskip
\bigskip
\bigskip

\Sectionvar Abstract

Inflation is known to be generically eternal to the future:
the false vacuum is thermalized in some regions of space,
while inflation continues in other regions.
Here, we address the question of whether inflation can also be
eternal to the past.  We argue that such a steady-state picture is
impossible and, therefore, that inflation must have had a beginning.
First, it is shown that the old inflationary model is not past-eternal.
Next, some necessary conditions are formulated for inflationary spacetimes
to be past-eternal and future-eternal. It is then shown that these
conditions cannot simultaneously hold in physically reasonable open
universes.

\Section Introduction

There are essentially two approaches to cosmology. The first, which may be
called evolutionary, assumes that the universe was born a
finite time ago. The task of cosmology is then to uncover
how the universe evolved from its initial state to its present form.
The second approach, which may be called the steady-state approach,
assumes that the universe has always been the way it is now. The task
of cosmology is then to understand the physical processes that sustain this
steady state.

An attractive feature of the steady-state approach is
that there is no need to impose any initial conditions on the universe
or to deal with the problem of the initial singularity. This approach
was extremely popular a few decades ago, but it fell out of favor in the
1960s under the pressure of observational evidence%
\Ref [See, for example, Peebles~P.~J.~E., \Jou{Principles of Physical
Cosmology\/} (Princeton University Press, Princeton, New Jersey (1993).].
Recently, however, steady-state cosmology has made a comeback
in the form of the eternal inflationary scenario.

Inflation is a state of rapid (quasi-exponential) expansion of the
universe%
\StoreRef{\Guth}%
\Refc [Guth A. H., \Jou{Phys. Rev. D\/} \Vol{23}, 347 (1981).]%
\StoreRef{\Sato}%
\Refm [Sato K., \Jou{Mon. Not. R. Astr. Soc.\/} \Vol{195}, 467 (1981).]%
\Refe [For reviews of inflation see, for example,
Blau S. K. and Guth A. H., in \Jou{300 Years of Gravitation},
edited by S. W. Hawking and W. Israel (Cambridge University Press,
Cambridge, England, 1987);\\
Linde A. D., \Jou{Particle Physics and Inflationary Cosmology\/}
(Harwood Academic, Chur Switzerland, 1990);\\
Kolb E. W. and Turner M. S., \Jou{The Early Universe\/}
(Addison-Wesley, New York, 1990).].
The inflationary expansion is driven by the potential energy of a
scalar field $\varphi$, while the field slowly ``rolls down'' its
potential $V(\varphi)$. When $\varphi$ reaches the minimum of the
potential, this vacuum energy thermalizes and inflation is followed by
the usual radiation-dominated expansion.

Soon after the inflationary scenario was proposed, it was realized
that, once started, inflation never ends completely%
\Refc [Vilenkin A., \Jou{Phys. Rev. D\/} {\bf 27}, 2848 (1983);\\
Steinhardt P. J., in \Jou{The Very Early Universe}, edited by
G. W. Gibbons, S. W. Hawking and S. T. C. Siklos (Cambridge University
Press, Cambridge, England, 1983);\\
Starobinsky A. A., in \Jou{Field Theory, Quantum Gravity and Strings},
Proceedings of the Seminar series, Meudon and Paris, France,
1984--1985, edited by M. J. de Vega and N. Sanchez, Lecture Notes in
Physics Vol. 246 (Springer-Verlag, New York, 1986);\\
Linde A. D., \Jou{Phys. Lett. B\/} {\bf 175}, 395 (1986).]%
\StoreRef{\AryaVi}%
\Refm [Aryal M. and Vilenkin A., \Jou{Phys. Lett. B\/}
{\bf 199}, 351 (1987).]%
\Refe [Goncharov A. S., Linde A. D. and Mukhanov V. F.,
\Jou{Int. J. Mod. Phys. A\/} {\bf 2}, 561 (1987);\\
Nakao K., Nambu Y. and Sasaki M., \Jou{Prog. Theor. Phys.\/}
{\bf 80}, 1041 (1988);\\
Linde A., Linde D. and Mezhlumian A., Stanford preprint SU-ITP-93-13 (1993).].
The evolution
of the field $\varphi$ is influenced by quantum fluctuations, and as
a result thermalization does not occur simultaneously in different parts
of the universe. Inflating regions constantly undergo thermalization,
but the exponential expansion of the remaining regions more than
compensates for the loss, and, at any time, there are parts of the universe
that are still inflating.

A model in which the inflationary phase has no end and continually
produces new islands of thermalization naturally leads to this question:
can this model also be extended to the infinite past, avoiding in this way
the problem of the initial singularity? The universe would then be
in a steady state of eternal inflation without a beginning.
We are going to argue here that the answer to this question is ``no''%
\StoreRef{\Vilenkin}%
\Refc [Vilenkin A., \Jou{Phys. Rev. D\/} \Vol{46}, 2355 (1992).]%
\StoreRef{\Borde}%
\Refm [Borde A., \Jou{Cl. and Quant. Gravity\/} \Vol{4}, 343 (1987).]%
\Refe [Borde A. and Vilenkin A., Tufts University cosmology preprint,
gr-qc 9312022 (1993).].

Before analyzing the general case, we clarify the ideas
using a simple model. In section~2 the possibility of eternal
inflation is discussed in the context of the ``old'' inflationary
scenario, in which the vacuum energy is strictly constant and vacuum
decay occurs through bubble nucleation. Old inflation is known to be
eternal to the future:
bubbles cannot fill the entire universe since the space between them is
expanding so fast~[\Guth,$\,$\Sato].
In the thermalized parts of the universe the distribution of matter
produced by colliding bubbles walls is grossly inhomogeneous, making old
inflation unsuitable as a realistic cosmological model. Here we
disregard this aspect of the problem and only concern ourselves with the
question of whether or not old inflation can be continued back to the
infinite past. We review the arguments showing that it cannot.

Using this discussion as a guide, we formulate in section~3
two conditions that
any spacetime describing an eternally inflating universe should
satisfy. (In the Appendix we give an example of spacetimes where these
conditions do not both hold, as well as an example where they do.)
We then show in section~4 that, under very general
assumptions, these conditions cannot be simultaneously satisfied.

We use the following conventions: the metric has signature
$(+, -, -, -)$ and Einstein's equation is
$R_{\mu\nu}-{1\over 2}g_{\mu\nu}R = 8\pi G\,T_{\mu\nu}$
(where $R_{\mu\nu}$ is the Ricci tensor associated with the metric
$g_{\mu\nu}$,
$R$ the scalar curvature and $T_{\mu\nu}$ the matter energy-momentum tensor).
Our main result uses the Penrose-Hawking-Geroch ``global techniques;''
an overview of the background that we need is given in section~4.
For further details and for the proofs of various standard global
results, see, for example, Hawking and Ellis\StoreRef{\HE}%
\Ref [Hawking S. W. and Ellis G. F. R., {\it The large scale structure of
spacetime\/} (Cambridge University Press, Cambridge, England, 1973).].

%----------------------------------------------------------------------------
\Section Old inflation

In the old inflationary scenario the false vacuum has the
energy-momentum tensor
$$
T_{\mu\nu} = \rho g_{\mu\nu}, \NumbEq
$$
where $\rho$ is constant. The homogeneous and isotropic solution of
Einstein's equation with $T_{\mu\nu}$ from \PrevEq\ is de~Sitter space,
which can be represented in the form
$$
ds^2 = dt^2 - e^{2Ht} d\vec{x}^2, \NumbEq
\NameEq{\deSittO}
$$
where
$$
H^2 = 8\pi G \rho / 3\ . \NumbEq
$$
This space has a horizon of radius~$H^{-1}$; observers separated by a
greater distance cannot communicate.

Bubbles nucleating in false vacuum expand, rapidly approaching the speed
of light. In de~Sitter space this corresponds to having asymptotically
static boundaries in co-moving coordinates. The physical radius of a bubble
formed at time~$t_1$ is (for $H(t-t_1) \gg 1$)
$$
r(t, t_1) \approx H^{-1} e^{H(t-t_1)}\ . \NumbEq
$$
An expanding bubble can affect the geometry of the outside region only
within a distance of approximately~$H^{-1}$ from its boundary. Hence,
although
bubbles carve large volumes out of de~Sitter space, the geometry of the
remaining regions is practically unchanged. We shall first assume that
inflation starts at some time~$t_0$ and later consider the limit
$t_0 \to -\infty$.

Bubble nucleation is a stochastic process with a constant
probability~$\lambda$ per unit spacetime volume. The probability for
no bubbles to be formed in a 4-volume~$\Omega$ is~[\Guth,%
\Refe [Guth A. H. and Weinberg E. J., \Jou{Phys. Rev. D\/} {\bf 23},
876 (1981).]
$$
{\cal P}_\Omega = e^{-\lambda\Omega}\ . \NumbEq
\NameEq{\Prob}
$$
Here is a quick derivation. Let ${\cal P}(A)$ be the probability for no
bubbles to nucleate in spacetime region~$A$. Then, for non-overlapping
regions~$A$ and~$B$, ${\cal P}(A\cup B) = {\cal P}(A) {\cal P}(B)$, and
for an infinitesimal volume~$d\Omega$, ${\cal P} \approx
1 - \lambda\,d\Omega$. The only function~${\cal P}(\Omega)$ with these
properties is~\PrevEq.

The probability for a given spacetime point $x=(t, \vec x)$ to be in
the inflationary phase is given by~\PrevEq\ with $\Omega$ being the volume
occupied by the false vacuum in the past light cone of~$x$. For
$H(t-t_0) \gg 1$, this volume is
$$
\Omega = {4\pi\over 3H^2}(t-t_0)\ . \NumbEq
$$
(This equation is easily understood if we note that null geodesics
continued to large negative values of~$t_0$ asymptotically approach the
horizon, which is a sphere of radius~$H^{-1}$.) From \PrevEqs1, \PrevEq,
the fraction of space that is still inflating at time~$t$ is
$$
f(t) \equiv \exp\left\{ -{4\pi\lambda \over 3H^3} (t-t_0) \right\}\ . \NumbEq
\NameEq{\Fract}
$$
The function~$f(t)$ decreases with time and vanishes as~$t \to +\infty$.
But the physical volume of the inflating regions,
$V(t) \propto e^{3Ht}f(t)$, grows with time. The reason is very simple:
for sufficiently small~$\lambda$, the rate of expansion of the false vacuum
regions is greater than their rate of decay.

\StartFigure
\strut\vbox to 145 bp{\hsize 216 bp\vss
\strut%
  }
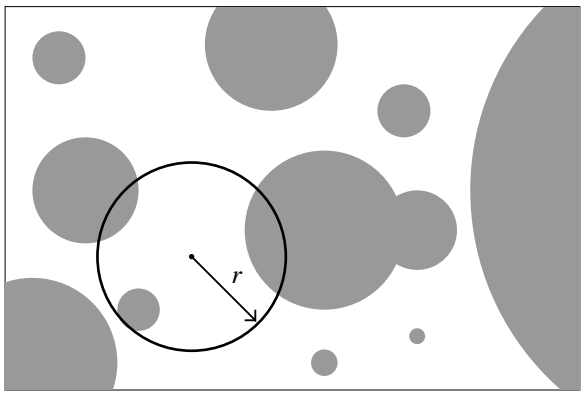
\caption{A schematic snapshot of the old inflationary universe. The
shaded regions represent bubbles and the white region represents the
inflationary background in which they are embedded.}
\EndFigure
An argument similar to that in Ref.~[\AryaVi]
shows that the inflating regions form a self-similar fractal of
dimension~\hbox{$d<3$}. This fractal dimension can be found from
$$
f(t) = \left({H^{-1}\over R}\right)^{3-d}, \NumbEq
$$
where~$H^{-1}$ is the size of the smallest bubbles and
$R\approx H^{-1} \exp[H(t-t_0)]$ is the size of the largest bubbles
(formed at $t\approx t_0$). A comparison of \PrevEqs1\ and \PrevEq\
gives
$$
d= 3 - {4\pi\lambda \over 3H^4}\ . \NumbEq
$$
\indent The meaning of the fractal dimension is easy to understand.
Consider a
sphere of radius~$r$ centered on a point in the inflating region
(see \Fig).
As~$r$ is increased, the volume~$V$ occupied by the false vacuum
inside the sphere grows (on average) proportional to~$r^4$. The deviation
of~$d$ from~$3$ can be attributed to the fact that, as the sphere becomes
larger, it is likely to include larger and larger bubbles. Of course, the
inflating regions have a fractal nature only on scales $H^{-1} < r < R$.
For $r > R$, $V\propto r^3$.

Let us now ask what happens if we remove the beginning of inflation to the
infinite past. As we said in the Introduction, we are
not concerned here with whether or not this model is realistic (it is
not). Rather, we are concerned with whether or not a consistent
model of eternal inflation is obtained by letting~$t_0 \to -\infty$.

As $t_0 \to -\infty$, the upper cutoff on the sizes of the bubbles is
removed; i.e., $R\to \infty$. At the same time the probability~\Prob\ for
a point to be in the inflationary phase and the fraction of space~$f$
occupied by the false vacuum both vanish. Note, however, that for a
point~$(\vec x, t)$ in an inflating region, there is a finite probability
that inflation will continue for any given time interval~$\Delta t$.
This probability is given by
$$
{\cal P}_{\Delta\Omega} = e^{-\lambda\Delta\Omega}, \NumbEq
\NameEq{\ProbTwo}
$$
where~$\Delta\Omega$ is the 4-volume between the past light cones
originating at~$(\vec x, t)$ and~$(\vec x, t+\Delta t)$; i.e.,
$$
\Delta\Omega = {4\pi\over 3H^3} \Delta t\ . \NumbEq
\NameEq{\Volu}
$$
The vanishing of~$f$ in~\Fract\ simply expresses the fact that an object
of fractal dimension~\hbox{$d<3$} cannot fill a 3-dimensional space; a
randomly chosen point is most likely to be inside an infinitely large
bubble. The physical volume occupied by the false vacuum is, however,
still increasing with time, and it may appear that we have a model of
eternal inflation.

The trouble with this model is that the metric~\deSittO\ is geodesically
incomplete~[\HE].
To see what this means, consider a flat spacetime
$$
ds^2 = dt^2 - dx^2 - dy^2 - dz^2 \NumbEq
$$
and introduce a new time coordinate~$\tau \equiv \ln t$. The metric
may be written in terms of the new time coordinate as
$$
ds^2 = e^{2\tau}d\tau^2 - dx^2 - dy^2 - dz^2 \NumbEq
$$
As~$\tau$ varies from~$-\infty$ to~$+\infty$, the Minkowski time~$t$
changes from~$0$ to~$\infty$, and thus the metric~\PrevEq\ covers
only half of Minkowski spacetime.

How could this have been determined without the prior knowledge
that~\PrevEq\ was obtained from~\PrevEqs1? One way to decide if there are
such ``missing regions'' to the past is to calculate the proper time along
past-directed timelike geodesics. For example, for the geodesic given
by $\vec x = \hbox{const}$, starting at~$\tau=0$ and continuing back
to~$\tau \to -\infty$, we have
$$
\int_{-\infty}^0 e^\tau d\tau = 1\ . \NumbEq
$$
The finiteness of this proper time indicates that our spacetime has an
``edge,'' and can perhaps be continued beyond $\tau=-\infty$.
(Note, however, that not all geodesically incomplete spacetimes can be
extended to complete spacetimes: if, for
instance, there are geodesics that run into a physical singularity, the
incompleteness is essential and cannot be removed.)
Similarly, we can check the finiteness of the affine parameter as
$\tau\to-\infty$ for null geodesics.

\StartFigure
\strut\vbox to 154 bp{\hsize 144 bp\vss
\strut%
  }
\label (142,155){$\biggl\uparrow\>\tilde t$}
\label (174,83){$\longleftarrow$ Not covered}
\label (30,17){$t=-\infty \longrightarrow$}
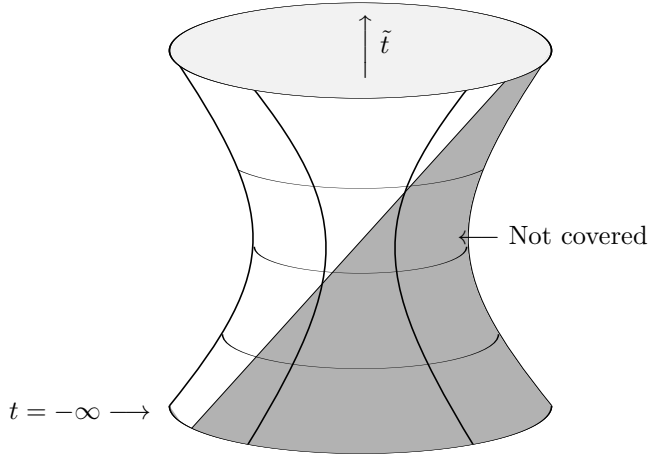
\caption{The full de Sitter spacetime. The time coordinate~$\tilde t$ points
upward and the spacelike cross sections orthogonal to~$\tilde t$
are 3-spheres. The shaded
portion of this spacetime lies to the past of the $t=-\infty$ surface
and is not covered by the old coordinates.}
\EndFigure
The metric~\deSittO\ which we used to describe the inflating universe
is both timelike and null geodesically incomplete. All timelike
geodesics, except the co-moving ones ($\vec x = \hbox{const}$), reach the
surface $t=-\infty$ in a finite proper time and all null geodesics
reach it within a finite lapse of their affine parameters. The full
de~Sitter space is covered by the metric~[\HE]
$$
ds^2 = d\tilde t^{\,2} - H^{-2} \cosh^2(H\tilde t)\, d\Omega_3^2 \NumbEq
\NameEq{\deSittT}
$$
where~$d\Omega_3^2$ is the metric on a unit 3-sphere (see \Fig).
In this extended
spacetime, the phase of the exponential expansion at~$\tilde t > 0$ is
preceded by a phase of exponential contraction at $\tilde t < 0$.
Of course, the contracting phase does not describe an inflating universe.
If such a contracting universe were filled by a false vacuum, the nucleating
bubbles would rapidly fill the space. The whole universe would
thermalize and collapse to a singularity without getting to the expanding
phase.

In terms of the probability~\ProbTwo\ for inflation to persist, a
nonzero answer was obtained only because the volume~$\Delta\Omega$
bounded by the two light cones was cut off at the surface~$t=-\infty$.
In the full de~Sitter space~\deSittT, $\Delta\Omega=\infty$ and
${\cal P}_{\Delta\Omega} = 0$.

%----------------------------------------------------------------------------
\Section Conditions for eternal inflation

The analysis in the previous section cannot be directly applied to realistic
inflationary scenarios. In realistic models the false vacuum energy~$\rho$
is replaced by the scalar field potential~$V(\varphi)$ which can vary in
space and time. The field~$\varphi$ is usually assumed to be slowly-varying
and the spacetime to be locally close to de~Sitter, but the global structure
of spacetime can be quite different. Moreover, quantum nucleation
of bubbles is replaced by a quantum random walk of the field~$\varphi$,
which is followed by thermalization when~$\varphi$ gets close to the bottom
of the potential. One can still find the probability for inflation to
persist at a given point for a specified period of time, but now this
probability depends on the initial value of~$\varphi$ at that point.
As a result, the locations of thermalization regions are strongly
correlated (unlike the bubble nucleation sites). The distribution
of thermalized regions obtained in a numerical simulation~[\AryaVi]
is shown in \NFig.
It can be shown that, as before, the inflating
region is a fractal whose dimension is determined by the shape of
the potential~$V(\varphi)$. Finally, if the magnitude of~$V(\varphi)$
gets near the Planck scale, the gravitational action may get significant
quantum corrections, and Einstein's equation can no longer be used.
\StartFigure
\noindent\vrule height 2 true in  depth 1 true in width0pt \hfil
\hbox{\Sixrm See figure~c in Ref.~[\AryaVi] for this
                     picture.}%
\caption{A computer simulation of the inflationary universe~[\AryaVi].
The dark areas represent thermalized regions
and the white ones the inflationary background.}
\EndFigure

In this section we shall formulate some necessary conditions that a
spacetime should satisfy in order to describe an eternally inflating
universe. We shall try to reduce to the minimum any assumptions about the
dynamical laws that govern the evolution of geometry and of the scalar
field~$\varphi$. However, to make the discussion meaningful we will have
to assume that, to a reasonable approximation, the spacetime can be
treated as a classical Riemannian manifold. Although eternal inflation
is sometimes described as occurring at the Planck scale, the description
invariably relies on classical spacetime concepts such as ``causality,''
``beginning,'' ``end,'' etc. It is also implicit in some of the
discussion that follows that spacetime
obeys the {\it causality condition\/}: i.e., it contains no closed null
or timelike curves.

In the previous section we saw that the spacetime~\deSittO\ describing
an inflating universe is geodesically incomplete. If eternal inflation is
possible, one should be able to construct a {\it complete\/} spacetime
that has the necessary properties of~\deSittO. This leads us to the
first of our conditions:

\smallskip\textindent{$\bullet$}
{\it Condition 1\/}: Timelike and null geodesics
are past-complete.
\smallskip

(We do not require future-completeness because that would preclude
the existence of
geodesics that encounter such things as black hole singularities.)

Now, it should be clear from the
preceding discussion that the essential property required for inflation
to be future-eternal is a non-zero probability for inflation to continue
at a given
point for a specified interval of time. In the old inflationary scenario
this probability is given by~\ProbTwo\ and its nonzero value is guaranteed
by the finiteness of the 4-volume~$\Delta\Omega$ in~\Volu.

To formulate the
corresponding requirement in the general case, we shall assume that the
boundaries of thermalized regions expand at a speed approaching the speed
of light, like the walls of bubbles expanding in a false vacuum. More
precisely, it will be assumed that a spacetime point~$p$ can be in an
inflating region only if its past, I$^-(p)$, intersects no thermalized
regions. In addition, it will be assumed that the probability of forming
thermalized regions does not vanish in the infinite past. Otherwise, it
is possible for the false vacuum to survive an infinitely long
contraction phase. This possibility, however, is against the spirit of
eternal inflation which assumes a ``steady-state'' picture of the universe.
With this assumption, the probability of having no thermalized regions in
an infinite volume vanishes, and we arrive at the following condition:

\smallskip \textindent{$\bullet$}
{\it Condition 2\/}: Let~$p$ and~$q$ be two points
in an inflating region with~$q$ to the future of~$p$. Then the
volume~$\Delta\Omega$ of the
difference of the pasts of~$q$ and~$p$ (i.e., the volume of
${\rm I}^-(q) - {\rm I}^-(p)$) is finite; i.e.,
$$
\Delta\Omega < \infty\ . \NumbEq
$$
\smallskip
The two conditions obtained here are very general. Neither rests on
detailed assumptions about inflation. In particular, we make no
assumptions about our spacetime being ``locally close to de Sitter.''
Such an assumption is often used to characterize inflationary scenarios.
A more general approach might be to require that the Ricci tensor obeys
$R_{\mu\nu}T^\mu T^\nu < 0$ for all timelike vectors (i.e.,
assuming Einstein's equation, that the
spacetime {\sl necessarily\/} violates the strong energy condition).
This, by itself, does not guarantee rapid expansion of the universe,
but it is necessary if such expansion is to take place. To see
this, observe that the geodesic focusing equation (the Landau-Raychaudhuri
equation) for a congruence of rotation-free timelike geodesics may be written
as~[\HE]
$$
\dot\theta \leq -{1\over 3}\theta^2 - R_{\mu\nu}T^\mu T^\nu, \NumbEq
$$
where the derivative is taken with respect to an affine parameter (the
usual choice is the proper time), $\theta=D_\mu T^\mu$ is the divergence
(or expansion) of the congruence, and~$T^\mu$ is the tangent field to the
geodesics with respect to the chosen parameter. If
$R_{\mu\nu}T^\mu T^\nu$ is positive at some point, then initially
parallel geodesics (i.e., $\theta=0$) that pass through that point will
start to converge (i.e., $\theta$ will become negative).
This is not the sort of behavior one expects in an inflating spacetime.

For the particular results of this paper, we do
not, however, need even such a general description of inflation.
Indeed, it is possible to ask if conditions~1 and~2 are likely to
hold in an arbitrary spacetime, whether it is inflating or not.
In the Appendix we show that both conditions cannot hold
in open Robertson-Walker spacetimes and we also construct an example
in which they do hold. This example is not realistic, however, for
in section~4 we show that a spacetime cannot simultaneously
satisfy our two conditions, as long as it meets
some mild (and physically reasonable) further requirements.

%----------------------------------------------------------------------------
\Section The general theorem

{\bf Theorem:}
{\sl A spacetime cannot simultaneously satisfy
the following conditions:
\item{A.} It is past causally simple.
\item{B.} It is open.
\item{C.} Einstein's equation holds, with a source that obeys the weak
  energy condition (i.e., the matter energy density is non-negative).
\item{D.} It is past null complete.
\item{E.} There is at least one point~$p$ such that for some
point~$q$ to the future of~$p$ the
volume of the difference of the pasts of~$q$ and~$p$ is finite.
\smallskip}

The proof of this theorem is supplied at the end of this section.
We discuss the assumptions of the theorem
in detail below, but here is a summary of what they mean:
Assumptions~A and~B are made solely for mathematical convenience.
The ultimate goal is to relax them (especially assumption~B).
Assumption~C holds in standard inflationary spacetimes, and is physically
quite reasonable.
Assumption~D is necessary for inflation to be past-eternal.
Assumption~E is necessary for inflation to be future-eternal.

Our result is similar in spirit to the standard singularity theorems
of general relativity: in all the singularity
results the existence of certain spacetime structures
(trapped surfaces, reconverging null cones, compact spacelike
hypersurfaces, etc.)
is shown to be incompatible with geodesic completeness.
None of the standard theorems, however, exactly fits the situation in
which we are interested.
For instance, several theorems (such as the multi-purpose Hawking-Penrose
theorem%
\StoreRef{\HP}%
\Ref [Hawking S. W. and Penrose R., \Jou{Proc. Roy. Soc. Lond.\/}
{\bf A314}, 529 (1970).])
assume the strong energy condition,
known to be violated in inflationary scenarios. (In fact, as we have
mentioned above, the violation of this energy condition is necessary
for inflation to occur.)
Others place much
stronger restrictions on the global causal structure of spacetime than
we do here (through assumption~A). More significantly, assumption~E
is entirely new~-- as we have seen above, it captures a characteristic
aspect of future-eternal inflationary spacetimes.

In order to discuss the assumptions in greater detail, we need certain
notations and results from global general relativity. A summary is given
below; the proofs of all our assertions may be found in Ref.~[{\HE}].

Let \cal M be a spacetime. (We assume that \cal M is time-orientable,
allowing a consistent global distinction between past and future.)
A curve in \cal M is called {\it causal\/} if it is everywhere
timelike or null (i.e., lightlike).
Let $p \in \cal M$. The
{\it causal\/} and {\it chronological\/} pasts of $p$, denoted respectively
by J$^-(p)$ and I$^-(p)$, are defined as follows:
\smallskip
J$^-(p) = \{q:$ there is a future-directed causal curve
           from $q$ to $p\}$,
\smallskip
\noindent and
\smallskip
I$^-(p) = \{q:$ there is a future-directed timelike curve
           from $q$ to $p\}$.
\smallskip
\noindent Thus J$^-(p)$ is the set of all points that can send signals
to~$p$ along timelike or null curves, and I$^-(p)$ is the set of all points
that can send signals to~$p$ along just timelike curves.

The {\it past light cone\/} of~$p$ may then be defined as E$^-(p) =
{\rm J}^-(p) - {\rm I}^-(p)$; i.e., E$^-(p)$ consists of all points
that can send signals to~$p$ along future-directed null curves, but not
along timelike curves. It may be shown
that the boundaries of the two kinds of pasts of~$p$ are the same; i.e.,
$\dot {\rm J}^-(p) = \dot {\rm I}^-(p)$.
Further, it may be shown that E$^-(p) \subset \dot {\rm I}^-(p)$.
In general, however, E$^-(p) \ne \dot {\rm I}^-(p)$; i.e, the
past light cone of $p$ (as we have defined it here) is a subset of the
boundary of the past of $p$, but is not necessarily the full boundary
of this past.
This is illustrated in \NFig.
\StartFigure
\strut\vbox to 115 bp{\hsize 190 bp \vss
\strut%
  }
\label(30,116){E$^-(p)$}
\label(220,105){$\dot {\rm I}^-(p)-{\rm E}^-(p) \ne \emptyset$}
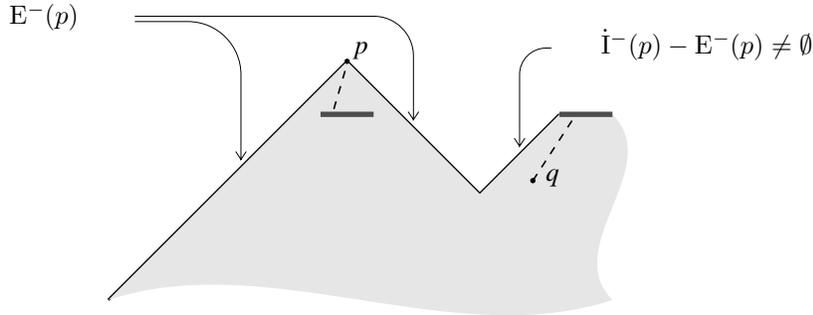
\caption{An example of the causal complications that can arise in an
unrestricted spacetime. Light rays travel along 45$^\circ$ lines in this
diagram, and the two thick horizontal lines are identified. This allows
the point~$q$ to send a signal to the point~$p$ along the dashed line,
as shown, even though~$q$ lies outside what is usually considered
the past light cone of~$p$. The boundary of the past of~$p$,
$\dot {\rm I}^-(p)$, then consists of the past light cone of~$p$,
$E^-(p)$, plus a further piece.
Such a spacetime is not ``causally simple.''}
\EndFigure

A set is called {\it achronal\/} if no two points in it can be connected by a
timelike curve; for instance, $\dot {\rm I}^-(p)$ is achronal
(if two points on it can be connected by a timelike
curve, then the pastmost of the two points will lie inside I$^-(p)$, not
on its boundary).

With this background information, we return to our discussion of
the assumptions of our theorem:

\smallskip\textindent{$\bullet$}
{\it Assumption A\/}: A spacetime is past causally simple
if E$^-(p) = \dot {\rm I}^-(p) \ne \emptyset$ for all points~$p$.
By saying that $\dot {\rm I}^-(p) \ne \emptyset$ we are ruling out causality
violations, and by saying that E$^-(p) = \dot {\rm I}^-(p)$ we are
excluding scenarios such as the one in \Fig.

In a causally simple space, the boundary of the past of a point~$p$ is
``generated by null geodesics to~$p$'': i.e.,
through each point $q\in \dot {\rm I}^-(p)$ there passes a future directed
null geodesic that has a future endpoint at~$p$~[\HE].

\smallskip\textindent{$\bullet$}
{\it Assumption B\/}: A universe is open if it contains no
compact achronal edgeless hypersurfaces. This is an extension
of the more common statement that a closed universe is one that contains
a compact spacelike hypersurface.

\smallskip\textindent{$\bullet$}
{\it Assumption C\/}:
An observer with four-velocity~$V^\mu$ will see a matter
energy density of $T_{\mu\nu}V^\mu V^\nu$. The weak energy condition is the
requirement that $T_{\mu\nu}V^\mu V^\nu \geq 0$ for all timelike
vectors~$V^\mu$.

This is a reasonable restriction on the matter fields in spacetime.
Indeed, the condition is known to be true for all classical matter.
If the matter fields are quantum fields, however, it is possible
for the weak energy condition to be violated by certain states of the
fields. Violations of the standard energy conditions have been discussed
by Tipler%
\Ref [Tipler F. J., \Jou{J. Diff. Eq.\/} {\bf 30}, 165 (1978);
\Jou{Phys. Rev. D\/} {\bf 17}, 2521 (1978).]
who has pointed out that such pointwise conditions may be replaced by weaker
integral, or averaged, conditions. An averaged weak energy condition has
been discussed by Roman%
\Ref [Roman T., \Jou{Phys. Rev. D\/} {\bf 33}, 3526 (1986);
\Jou{Phys. Rev. D\/} {\bf 37}, 546 (1988).],
and even weaker integral conditions have also been introduced~[\Borde].

It follows by continuity from the weak energy condition that
$T_{\mu\nu}N^\mu N^\nu \geq 0$ for all null vectors $N^\mu$ and from
Einstein's
equation that $R_{\mu\nu}N^\mu N^\nu \geq 0$. It is this final form, sometimes
called the {\it null convergence condition}, that we will actually use.
(Thus our results will remain true even in other theories of gravity, as long
as the null convergence condition, or one of the weaker corresponding
integral conditions, continues to hold.)

\smallskip\textindent{$\bullet$}
{\it Assumptions D \& E\/}: These
have been discussed in detail in the previous section.
\smallskip

Here, now, is the proof of our main result:

\smallskip \noindent {\bf Proof:}
Suppose that a spacetime obeys assumptions~A--E.
We show in two steps that a contradiction ensues.

\textindent{1.} {\it A point $p$ that satisfies assumption~E has a finite
past light cone.} By a ``finite past light cone'' is meant a light cone
E$^-(p)$ such that every past-directed null geodesic that initially
lies in the cone leaves it a finite affine parameter distance to the
past of~$p$.

Suppose, to the contrary, that a null geodesic $\gamma$ lies in E$^-(p)$
an infinite affine parameter distance to the past of~$p$. Let~$v$ be
an affine parameter on~$\gamma$ chosen to increase to the past
and to have the value~0 at~$p$. Consider a small `conical' pencil
of null geodesics in E$^-(p)$ around~$\gamma$ and choose coordinates
$x^1$ and~$x^2$ on the spacelike cross sections of this pencil.
Now vary this pencil along some timelike curve between~$p$ and~$q$.
This sets up a null geodesic congruence;
let $N^\mu$ be the tangent vector field to these geodesics associated with
the affine parameter~$v$.
If~$u$ is a parameter along the timelike curve between~$p$ and~$q$,
the surfaces of constant~$u$
will each consist of a pencil of null geodesics; choose~$u$ so that
the pencil containing~$\gamma$ corresponds to~$u=0$. See \NFig\ for
an illustration of this construction.
\StartFigure
\strut\vbox to 125 bp{\hsize 60 bp \vss
\strut%
}
\label(111,120){$\bigl\uparrow$ to\ \ $q$}
\label(103,60){$p$}
\label(80,48){$(u=0)$}
\label(74,110){$(u=\Delta)$}
\label(110,84){$\Delta$}
\label(107,22){${\cal A}(v)$}
\label(173,8){$\gamma$}
\label(153,19){$\searrow N^\mu$}
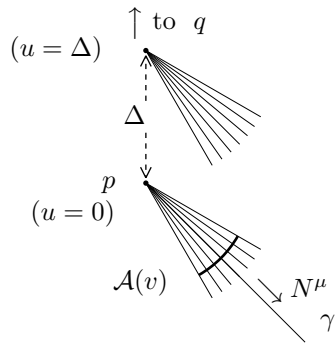
\caption{The geodesic congruence and the volume of interest. A small
pencil of null geodesics emanating in the past direction from~$p$
is chosen around the geodesic of interest,~$\gamma$. The pencil is
varied in the future direction till the vertex is at the point~$q$ (not
shown above).}
\EndFigure

How do the null geodesics in this
congruence move away from or towards~$\gamma$? If one considers
a ``deviation vector'' $Z^\mu$ that connects points on~$\gamma$ to points on
a nearby geodesic~[{\HE}], a straightforward calculation
yields the result that the only physically relevant variations
in~$Z^\mu$ come from its components in the spacelike 2-space
in E$^-(p)$ that is orthogonal~to~$N^\mu$. This is most easily
seen by introducing a pseudo-%
orthonormal basis $\{N^\mu, L^\nu, X^\mu_1, X^\mu_2\}$
where $L^\mu$ is a null vector
such that $N^\mu L_\mu = 1$ and $X^\mu_1$ and $X^\mu_2$
are unit spacelike vectors orthogonal to $N^\mu$ and~$L^\mu$.
The deviation vector~$Z^\mu$ may be written
as $Z^\mu = n N^\mu + l L^\mu + x_1 X^\mu_1 + x_2 X^\mu_2$.
Now, it is possible to choose the affine parametrization so that $n=0$.
Further, $l=Z^\mu N_\mu$ and the derivative
of $l$ vanishes in the direction of $N^\mu$
(i.e., $N^\nu D_\nu(Z^\mu N_\mu)=0$).

The metric may be expressed in terms of the coordinates $(u, v, x^1, x^2)$
as~[\Vilenkin]
$$
ds^2=g_{uu}du^2 + 2\,du\,dv +2g_{ui}du\,dx^i + g_{ij}dx^idx^j, \NumbEq
$$
where~$i$ and~$j$ run from~$1$ to~$2$.
The determinant of this metric is~$g=-^{(2)\!}g$, where~$^{(2)\!}g$ is the
determinant of~$g_{ij}$.

All of this means that the volume of the spacetime region occupied by the
portion of the geodesic congruence between~$u=0$ and~$u=\Delta$
(where~$\Delta$ is infinitesimal) may be expressed as
$$
\Delta V = \Delta \int_0^\infty {\cal A}(v)\, dv, \NumbEq
$$
where
$$
{\cal A}(v) = \int \sqrt{^{(2)\!}g}\, d^2x \NumbEq
$$
is the area of the spacelike cross section
of the light cone orthogonal to~$N^\mu$ (see \Fig).
The region whose volume we are calculating is a subset of
$\overline{{\rm I}^-(q) - {\rm I}^-(p)}$ (i.e., the closure of
the difference of the pasts of~$q$ and~$p$)
and it must thus have a finite volume.
In order for this to happen, ${\cal A}$ must
decrease somewhere along~$\gamma$.

The propagation equation for $\cal A$ is~[{\HE}]
$$
\dot{\cal A} = \theta {\cal A} \NumbEq
$$
where a dot represents a derivative with respect to~$v$ and $\theta =
D_\mu N^\mu$ is the divergence of the congruence. If ${\cal A}$ decreases
it follows that $\theta$ must become negative. But the propagation
equation for $\theta$ may be written as~[{\HE}]
$$
\dot\theta \leq -{1\over 2}\theta^2 - R_{\mu\nu}N^\mu N^\nu
              \leq -{1\over 2}\theta^2 \NumbEq
$$
(where we have used assumption~C in the last step).
If $\theta < 0 $ somewhere, it
follows that $\theta \to -\infty$ within a finite affine parameter distance.

The divergence of $\theta$ to $-\infty$ is a signal that the null
geodesics from~$p$ have refocused. It is a standard result in global
general relativity~[{\HE}] that points on such null geodesics beyond
the focal point enter the interior of the past light cone
(i.e., enter I$^-(p)$) and no longer lie in E$^-(p)$.
This is illustrated in \NFig.
\StartFigure
\strut\vbox to 110 bp{\hsize 160 bp \vss
\vskip15pt\strut%
}
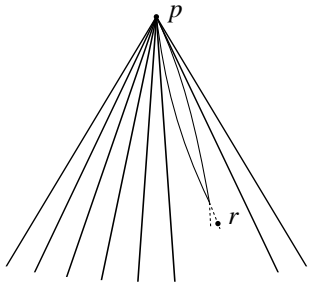
\caption{The past light cone of~$p$, showing (in an exaggerated way) the
focusing of some null geodesics. Points~$r$ beyond the focal point
lie in the interior of the past of~$p$ (i.e., in I$^-(p)$) and no longer
on the boundary of the past. The thrust of step~1 of the proof is to
show that such focusing occurs along every past-directed null geodesic
from~$p$ that lies on the boundary (i.e., on~E$^-(p)$), for a sufficient
affine length.}
\EndFigure

Thus the null cone E$^-(p)$ must be finite in the sense defined above.

\textindent{2.} {\it The result of step 1 contradicts assumption B.}
{}From causal simplicity it follows that E$^-(p)$ (being equal to the full
boundary of the past of $p$, $\dot{\rm I}^-(p)$) is an edgeless surface.
It is also achronal. And step~1 implies that E$^-(p)$
is compact. These three statements taken together contradict assumption~B.

\Section Discussion

The argument given above shows that inflation does
not seem to avoid the problem of the initial singularity (although it
does move it back into an indefinite past). In fact, our analysis
of assumption~E in section~3 suggests that almost all points in the inflating
region will have a singularity somewhere in their pasts.
In this sense, our result is stronger than most of the usual singularity
results (which, in general, predict the existence of just one
singularity, and in the case of the Hawking-Penrose theorem~[\HP] fail to
provide any information at all about the location of this singularity).

The only way to deal with the problem of the initial singularity
is probably to treat the universe quantum mechanically and
describe it by a wave function rather than by a classical spacetime.

The theorem that we have proved here is based on several
assumptions which it would be desirable to further justify or relax.
The principal relaxation that is necessary is in assumption~B;
i.e., closed universes must also be accommodated. It may appear at first
sight that this will be difficult to achieve since assumption~B entered
into the proof above at a crucial place. However, all that we really
need is to exclude situations where null geodesics recross after running
around the whole universe (as they do in the static Einstein model).
If the reconvergence of the null cone discussed above
occurs on a scale smaller than the cosmological one, then essentially the
same argument goes through.  This approach to applying open universe
singularity theorems to closed universes has been outlined previously
by Penrose\Ref [Penrose R., in {\it Battelle Rencontres}, edited by
C. M. DeWitt and J. A. Wheeler (W. A. Benjamin, New York, 1968).]
and it will be discussed in detail separately as will the relaxation
of assumption~A.

Assumption~E, which plays a central role in our argument, requires
further justification. It rests on the assumptions that (i)~the boundaries
of thermalized regions expand at speeds approaching the speed of light,
and (ii)~that the probability of finding no thermalized regions in an
infinite spacetime volume vanishes. Both assumptions are plausible
and, as we have seen in section~2, they are true in the original
inflationary scenario.
It would be interesting, however, to determine the exact conditions of
validity for these assumptions and to investigate the possibility
of relaxing them.

%----------------------------------------------------------------------------
\AppndEq
\Sectionvar Appendix: Geodesic completeness and past volumes

Consider the open Robertson-Walker metrics; they may be expressed
in the form
$$
ds^2 = a^2(\eta) \bigl[d\eta^2 - d\chi^2 - f^2(\chi)\,d\Omega^2 \bigr],
\NumbEq
$$
where $d\Omega^2=d\theta^2 + \sin^2\theta\,d\phi^2$ is the metric on a unit
2-sphere, $\chi \geq 0$, and
$$
f(\chi) = \cases{\chi&\quad(spatially flat open universes)\cr
                 \sinh\chi&\quad(spatially curved open universes)\cr}\ .
\NumbEq
$$
The range of the coordinate~$\eta$ depends
on the behavior of the function~$a(\eta)$. As we shall see below, the range
may or may not be infinite to the past.

The null geodesics that pass through $\chi=0$ are given by
$$
\eqalignno{
a^2(\eta){d\eta\over dv}&= \hbox{constant},& \numbeq\cr
\noalign{\noindent and}
\chi(v)&=\pm\eta(v) + \hbox{constant},& \numbeq\cr}
$$
where~$v$ is an affine parameter.
If the geodesics are to be past complete, $v$ must approach
$-\infty$ to the past (assuming that it is
chosen to decrease in the past direction).
The time coordinate~$\eta$ need not itself approach~$-\infty$:
from \PrevEqs1\ it follows that
all that is needed in order to guarantee past
completeness for these null geodesics is that
$$
\int_{\eta_{\rm min}}^\eta \!\! a^2(\hat\eta)\,d\hat\eta \NumbEq
\NameEq{\GeoComp}
$$
diverge (where $\eta_{\rm min}$ represents the lower bound on $\eta$).
If $\eta_{\rm min}$ is finite, the divergence of \PrevEq\ must arise from
a blowing up of the function~$a^2(\eta)$ as $\eta\to\eta_{\rm min}$.
In this case, the universe starts off ``infinitely stretched'' and, initially
at least, contracts. Such a picture is inconsistent with inflation.
But for the discussion below, it is not necessary to exclude this case
and we make no assumptions here about whether or not~$\eta_{\rm min}$
is finite.

If~$\eta$ is bounded below by~\hbox{$\eta_{\rm min}$}, then~$\chi$ will
be bounded above along each null geodesic by some value~$\chi_{\rm max}$
(finite if $\eta_{\rm min}$ is, and infinite otherwise)
when the geodesic is followed into the past.

Consider the volume~$\Delta\Omega$ between the past light cones
of $(\chi=0, \eta=0)$ and $(\chi=0, \eta=\Delta)$. Since
$\sqrt{-g}= a^4(\eta)f^2(\chi)\sin\theta$, we have
$$
\Delta\Omega = \int_0^{\chi_{\rm max}} \!\! d\chi
               \int_{-\chi}^{-\chi+\Delta} \!\! d\eta
               \int_0^\pi \!\! d\theta
               \int_0^{2\pi} \!\! d\phi\> a^4(\eta) f^2(\chi) \sin\theta\ .
               \NumbEq
$$
For small $\Delta$ this becomes
$$
\Delta\Omega\approx 4\pi\Delta\int_0^{\chi_{\rm max}}
\!\! d\chi\> a^4(-\chi) f^2(\chi)\ .\NumbEq
$$
Now, the divergence of \GeoComp, necessary for past
null completeness, implies that
$$
\int_0^{\chi_{\rm max}}\!\! d\chi\> a^2(-\chi) \NumbEq
$$
diverges. Thus $\int\! a^4\,d\chi$ must also diverge.
Further,
$f^2$ is an increasing function. (Actually, all that we really need is that
$f^2$ remain greater than some positive constant
as $\chi\to\chi_{\rm max}$.)
The conclusion from these observations is that~$\Delta\Omega$ in \PrevEqs1\
diverges to~$+\infty$ in
the open Robertson-Walker models if they are past null complete.

Thus past geodesic
completeness is incompatible here with~$\Delta\Omega < \infty$.
It might seem that this must always be so, no matter what the metric.
(I.e., it might seem that the spacetime volume between past-complete
null cones must necessarily be infinite.)
We now present an example showing that this is not the case.
Consider the metric
$$
ds^2 = A^2(\eta)(d\eta^2 - d\chi^2) - B^2(\eta) \chi^2 d\Omega^2\ . \NumbEq
$$
If the null geodesics that emanate from $(\eta=0, \chi=0)$ are to be past
complete, and the coordinate~$\eta$ is to approach~$-\infty$ when these
geodesics are followed into the past, we must have
$$
\int_{-\infty}^0 \!\! A^2(\eta)\, d\eta = \infty\ . \NumbEq
$$
The past volume difference that we are interested in is now
$$
\eqalignno{
\Delta\Omega&\approx 4\pi\Delta \int_0^{\infty} \!\! d\chi\,
\chi^2 A^2(-\chi) B^2(-\chi)&\cr
&= 4\pi\Delta \int_{-\infty}^0 \!\! d\eta\,
\eta^2 A^2(\eta) B^2(\eta)\ .&\numbeq\cr}
$$
This integral will converge (and \PrevEqs1\ will diverge) if, for instance,
$$
\eqalignno{
A(\eta)&=(1-\eta)^{-\alpha}, \quad 0 \leq \alpha < {1\over 2}, &\numbeq\cr
\noalign{\noindent and}
B(\eta)&=(1-\eta)^{-\beta}, \quad \beta > 1\ .&\numbeq\cr}
$$

This example escapes the clutches of the general theorem of section~4
because it violates the weak energy condition. It is instructive to examine
how exactly the escape occurs. Let~$p$ be a point on the $\chi=0$ line with
$\eta=\eta_0 < 1$.
The divergence of the past-directed null
geodesics from~$p$ is
$$
\theta = {2\over A^2}\left({1\over\eta_0-\eta} -
{1\over B}{dB\over d\eta}\right)\ .
\NumbEq
$$
If~$A$ and $B$ are given by \PrevEqs2\ and \PrevEqs1,
$\theta$ diverges to~$+\infty$ at~$p$ and it becomes negative for
$$
\eta<{\beta\eta_0 - 1\over \beta - 1}\ . \NumbEq
$$
This means that the null geodesics from~$p$ have started to reconverge.
This reconvergence is necessary for a finite
past-volume, as is shown in the main argument in section~4.
But the violation of the weak energy condition
here allows the geodesics to avoid refocusing
(an examination of \PrevEqs1\ reveals that it remains finite everywhere
to the past of~$p$), thereby providing an escape from our theorem.

%----------------------------------------------------------------------------
\Sectionvar Acknowledgements

One of the authors~(A.V.) thanks the organizers of the
Eighth Nishinomiya Yukawa Symposium for their warm hospitality.
He also acknowledges partial support from the
National Science Foundation. The other author~(A.B.) thanks the Institute
of Cosmology at Tufts University for its hospitality over the period
when this work was done.

%----------------------------------------------------------------------------
\ShowReferences

\bye